\documentclass[a4paper,11pt]{article}
\usepackage[dvips]{graphicx}
\usepackage{amsmath}
\usepackage{amssymb}
\usepackage{epsfig}
\usepackage{float}

\title{Phase Structure of a Four- and Eight-Fermion Interaction Model at Finite Temperature and Chemical Potential in Arbitrary Dimensions}

\author{{\normalsize Masako Hayashi}\\
{\small {\it Department of Physics, Hiroshima University,}}\\
{\small {\it Higashi-Hiroshima 739-8526, Japan}}\\[2mm]
{\normalsize Tomohiro Inagaki}\\
{\small {\it Information Media Center, Hiroshima University,}}\\
{\small {\it Higashi-Hiroshima 739-8521, Japan}}\\[2mm]
{\normalsize Wataru Sakamoto}\\
{\small {\it Graduate School of Integrated Arts and Sciences, Hiroshima University,}}\\
{\small {\it Higashi-Hiroshima, 739-8521, Japan}}
}

\date{}

\begin{document}

\maketitle

\begin{abstract}
The phase structure of a four- and eight-fermion interaction model is investigated at finite temperature and chemical potential in arbitrary space-time dimensions, $2\leq D<4$. The effective potential and the gap equation are calculated in the leading order of the $1/N$ expansion. If the first order phase transition takes place, the phase boundary dividing the symmetric and the broken phase is modified by the eight-fermion interaction.
\end{abstract}

\section{Introduction}	
Models of particle physics can be classified based on its symmetry. Some apparent symmetry in the models is broken spontaneously. The basic ingredient of the standard electroweak theory is the spontaneous symmetry breaking of the $SU(2)\otimes U(1)$ gauge symmetry and grand unified theories are constructed on the basis of the spontaneous breaking of higher gauge symmetry at high energy scale. The phase transition has played a decisive role in the scenario of cosmological and astrophysical evolution. There is a possibility to inspect models of particle physics in cosmological and astrophysical objects. It is quite interesting to make investigations on the spontaneous symmetry breaking under the circumstance of the early universe, high temperature, chemical potential and so on.

As is known, a strong gauge interaction between fermions is subject to chiral symmetry breaking in a dynamical way at vanishing temperature and chemical potential. The broken symmetry is restored at high temperature and chemical potential. The symmetry breaking takes place in the strong coupling region in the gauge field theories. Non-perturbative treatment is necessary to study its phase structure. We often use phenomenological effective models to evaluate non-perturbative effects for the gauge theories. A four-fermion interaction model is one of the simplest models where the chiral symmetry is broken dynamically \cite{NJL,GN}. It is also useful to describe meson properties as a low energy effective theory of QCD. 

In two space-time dimensions four-fermion interactions are given by marginal operators. Hence the model is renormalizable. In three space-time dimensions the model is solvable and renormalizable in the sense of the $1/N$ expansion. We can also regard the model in the space-time dimensions less than four as a regularized model in four dimensions \cite{Dim,Dim2}. Many works have been done to study the phase structure of a four-fermion interaction model at finite temperature and chemical potential in a space-time dimension less than four. The critical temperature in the Gross-Neveu model has been found at the large $N$ limit in Ref.~\cite{CT}. The influence of chemical potential is discussed in Ref.~\cite{FC}. The phase structure at finite temperature and chemical potential has been studied in two \cite{Tr,Wol,Klimenko:1986uq}, three \cite{Kli}, four \cite{Asakawa:1989bq} and an arbitrary space-time dimensions \cite{Dim2,IKM,Zhou:2002fy}.

It should be noticed that there is no reason to omit higher dimensional operators in a phenomenological effective model at high temperature and chemical potential. For example, 't~Hooft consider a six fermion interaction in a low energy effective model in QCD to deal with the axial anomaly \cite{tHooft}. An eight-fermion interaction has been discussed to describe the mass spectrum for the nonet pseudoscalar meson \cite{Alkofer:1990uh,OHBP}. Higher derivative interaction has been studied in Ref.~\cite{ELO3}. These interactions are given by irrelevant operators. The renormalized coupling constant should be divergent at a high energy scale. Thus a larger contribution is expected from irrelevant operators at high temperature and chemical potential \cite{Hiller:2008nu,Moreira:2010in}.

In the present paper we consider scalar type four- and eight-fermion interactions and investigate the contribution of the eight-fermion interaction to the phase structure at finite temperature and chemical potential in arbitrary space-time dimensions, $2\leq D <4$. In Sec.~2 we define a model with four- and eight-fermion interactions considered here. The effective potential is calculated in the leading order of the $1/N$ expansion. In Sec.~3 we introduce the temperature and chemical potential in the model according with the imaginary time formalism. In Sec.~4 we numerically evaluate the behavior of the effective potential with as the temperature and chemical potential vary. Observing the minimum of the effective potential, we derive the phase boundary of the discrete chiral symmetry. In Sec.~5 we give some concluding remarks,

\section{Four- and Eight-Fermion Interaction Model}
Here we consider a model with only scalar type four- and eight-fermion interactions to find a characteristic feature of the contribution from a higher dimensional operator to dynamical symmetry breaking at finite temperature and chemical potential. We extend the Gross-Neveu type model to have a scalar type eight-fermion interaction in arbitrary space-time dimensions, $2\leq D <4$ and start with the action \cite{HIT},
\begin{eqnarray}
S &=&\int d^{D}x  \ \left[ \sum^{N}_{i=1} \bar{\psi_{i}}(x) i \gamma^{\mu} \partial_{\mu} \psi_{i}(x)  \right. \nonumber \\
 && \left. \ \ \ + \frac{G_{1}}{N} \left( \sum^{N}_{i=1} \bar{\psi_{i}}(x) \psi_{i}(x)\right)^2 + \frac{G_{2}}{N^{3}} \left( \sum^{N}_{i} \bar{\psi_{i}}(x) \psi_{i}(x) \right)^4 \right] ,
\label{act:org}
\end{eqnarray}
where the lower index $i$ represents the flavors of the fermion field $\psi$, $N$ is the number of the fermion species and $G_1$, $G_2$ are coupling constants. Below we neglect the flavor index for simplicity. The action (\ref{act:org}) is invariant under the discrete chiral transformation,
\begin{equation}
\bar{\psi}(x) \psi(x) \rightarrow -\bar{\psi}(x) \psi(x) .
\end{equation}
This chiral symmetry prohibits the action to have a mass term. If the composite operator, $\bar{\psi}(x) \psi(x)$, develops a non-vanishing expectation value, above chiral symmetry is eventually broken.

In practical calculations it is more convenient to introduce auxiliary fields, $s$ and $\sigma$, and rewrite the action (\ref{act:org}) with the equivalent one \cite{Reinhardt:1988xu},
\begin{eqnarray}
S_y &=&\int d^{D}x  \ \left[ \bar{\psi}(x) (i \gamma^{\mu} \partial_{\mu}-s)  \psi(x)
+ \frac{N}{4G_{1}} \sigma^2 - \frac{N}{2G_{1}} s \sigma
+\frac{NG_2}{16{G_{1}}^4} \sigma^4 \right] .
\label{act:aux}
\end{eqnarray}
Replacing the auxiliary fields in the action (\ref{act:aux}) by the solution of the equations of motion,
\begin{eqnarray}
\sigma = -\frac{2G_{1}}{N} \bar{\psi} \psi ,
\end{eqnarray}
and
\begin{eqnarray}
s= \sigma +\frac{G_{2}}{2G_{1}}\sigma^{3} ,
\label{def:s}
\end{eqnarray}
we reproduce the original action (\ref{act:org}). The expectation value for the composite operator, $\bar{\psi} \psi$ is measured by the one for the auxiliary field $\sigma$. Thus we regard the expectation value, $\langle\sigma\rangle$, as an order parameter for the chiral symmetry breaking. If the non-vanishing expectation value is assigned to $s$, there appears a mass term for the fermion field, $\psi$. The dynamically generated fermion mass, $m_d\equiv \langle s\rangle$, is given by a function of $\langle\sigma\rangle$.

We want to calculate the order parameter, $\langle\sigma\rangle$, under the ground state. It is found by observing the minimum of the effective potential. In the leading order of the $1/N$ expansion the effective potential is written as \cite{IMO,IH,HI}
\begin{eqnarray}
V(s,\sigma)=-\frac{1}{4G_{1}}\sigma^2+\frac{1}{2G_{1}}s \sigma -\frac{G_{2}}{16G_{1}^{4}}\sigma^{4}-i \mathrm{tr} \int_{0}^{s}dm S(x,x;m) ,
\label{EP}
\end{eqnarray}
where $S(x,x;m)$ is the spinor two-point function and $\mathrm{tr}$ denotes trace with respect to spinor index. This expression is also applicable to the state in thermal equilibrium.  For vanishing temperature and chemical potential the two-point function is given by
\begin{eqnarray}
S(x,y;s)=\int \frac{d^{D}k}{(2 \pi)^{D}}e^{-ik(x-y)}\frac{1}{\gamma^{\mu}k_{\mu}-s} .
\label{Spi2}
\end{eqnarray}
Substituting Eq.(\ref{Spi2}) to Eq.(\ref{EP}) and performing the integration, we obtain the effective potential for vanishing temperature and chemical potential,
\begin{eqnarray}
V_{0}(s, \sigma) = -\frac{1}{4G_{1}} \sigma^{2} + \frac{1}{2G_{1}} s \sigma - \frac{G_{2}}{16{G_{1}}^{4}} \sigma^4 
\nonumber \\ 
- \frac{\mathrm{tr}1}{ \left( 4 \pi \right)^{D/2} D} \Gamma \left( 1- \frac{D}{2} \right) \left( s^{2} \right)^{D/2} ,
\label{v0}
\end{eqnarray}
where we put the suffix $0$ for $V_{0}(s,\sigma)$ to keep the memory that $T=\mu=0$.

On the ground state the expectation value of the auxiliary fields, $s$ and $\sigma$, should satisfy the stationary conditions,
\begin{eqnarray}
\left.\frac{\partial V_0}{\partial s}\right|_{\sigma}=
\frac{1}{2G_1}\sigma-\frac{\mathrm{tr}1}{ \left( 4 \pi \right)^{D/2}} \Gamma \left( 1- \frac{D}{2} \right) s \left( s^{2} \right)^{D/2-1} =0 ,
\label{stationary1}
\end{eqnarray}
and
\begin{eqnarray}
\left.\frac{\partial V_0}{\partial \sigma}\right|_{s}=
-\frac{1}{2G_1}\sigma +\frac{1}{2G_1}s-\frac{G_2}{4{G_1}^4}\sigma^3
 =0 .
\label{stationary2}
\end{eqnarray}
Thus the auxiliary field, $s$, is described by a function of $\sigma$,
\begin{eqnarray}
s=\sigma + \frac{G_2}{2{G_1}^3}\sigma^3 .
\label{s1}
\end{eqnarray}
Using Eqs.(\ref{stationary1}) and (\ref{s1}), we obtain the self-consistency equation for the dynamically generated fermion mass,
\begin{eqnarray}
s&=&
2 G_1 \frac{\mathrm{tr}1}{ \left( 4 \pi \right)^{D/2}} 
\Gamma \left( 1- \frac{D}{2} \right) s \left( s^{2} \right)^{D/2-1}
\nonumber \\
&&+ 4G_2\left[\frac{\mathrm{tr}1}{ \left( 4 \pi \right)^{D/2}} \Gamma \left( 1- \frac{D}{2} \right) s \left( s^{2} \right)^{D/2-1}\right]^3 .
\label{s2}
\end{eqnarray}
This equation has a trivial solution $s=0$. Since $\Gamma \left( 1- D/2 \right)$ is negative for $2\leq D<4$, Eq(\ref{s2}) has a real and non-trivial solution only if either $G_1$ or $G_2$ is negative. In the four-fermion interaction model, $G_2=0$, the discrete chiral symmetry is broken for a negative $G_1$ \cite{IKM}. The solution is given by
\begin{eqnarray}
s=m_{0}\equiv \left[ 2 G_{1}\frac{ \mathrm{tr}1}{ \left( 4 \pi \right)^{D/2}} \Gamma \left( 1- \frac{D}{2} \right) \right]^{1/(2-D)} .
\end{eqnarray}

On the stationary condition Eq.(\ref{stationary2}), the effective potential (\ref{v0}) is written as a function of the auxiliary field, $\sigma$,
\begin{eqnarray}
\frac{V_{0}(\sigma)}{m_{0}^{D}} &=& \frac{\mathrm{tr}1}{ (4 \pi)^{D/2}} \Gamma \left(  1- \frac{D}{2} \right)
\nonumber \\
&& \times \left\{ \frac{1}{2} \left( \frac{\sigma}{m_{0}} \right)^{2} + \frac{3g}{4} \left( \frac{\sigma}{m_{0}} \right)^{4} - \frac{1}{D} \left( \left[ \frac{\sigma}{m_{0}} + g \left( \frac{\sigma}{m_{0}} \right)^{3} \right]^{2} \right)^{D/2} \right\} ,
\nonumber\\
\label{v0:2}
\end{eqnarray}
where we normalize all the mass scale by $m_0$ and set
\begin{eqnarray}
g\equiv\frac{G_{2} m_{0}^{2}}{2 G_{1}^{3}} .
\end{eqnarray}

We can regard the expectation value, $\langle\sigma\rangle$, as the order parameter of the chiral symmetry breaking. On the ground state it is given by the minimum of the effective potential (\ref{v0:2}). The extremum of the effective potential satisfies the stationary condition,
\begin{eqnarray}
&&\frac{1}{{m_0}^{D-1}}\frac{\partial V_0(\sigma)}{\partial \sigma}
\nonumber \\
&&\ \ = \frac{\mathrm{tr}1}{ (4 \pi)^{D/2}} \Gamma \left(  1- \frac{D}{2} \right) \left(1+3g\left( \frac{\sigma}{m_{0}} \right)^{2}\right)
\nonumber \\
&& \ \ \ \ \ \ \times 
 \left\{ \frac{\sigma}{m_{0}}
  -  \left[ \frac{\sigma}{m_{0}} + 
   g \left( \frac{\sigma}{m_{0}} \right)^{3} \right]
\left( \left[ \frac{\sigma}{m_{0}} + g \left( \frac{\sigma}{m_{0}} \right)^{3} \right]^{2} \right)^{D/2-1} 
\right\} = 0 .
\nonumber\\
\label{stationary3}
\end{eqnarray}
This equation has a trivial solution,
\begin{eqnarray}
\frac{\sigma}{m_{0}}= 0,
\label{sol0}
\end{eqnarray}
and non-trivial ones,
\begin{eqnarray}
\left( \frac{\sigma}{m_{0}} \right)^{2}=-\frac{1}{3g}
=-\frac{2G_1^3}{3G_2m_0^2} ,
\label{sol1}
\end{eqnarray}
and
\begin{eqnarray}
1= 
\left(1+g\left( \frac{\sigma}{m_{0}} \right)^{2}\right)
\left( \left[ \frac{\sigma}{m_{0}} + g \left( \frac{\sigma}{m_{0}} \right)^{3} \right]^{2} \right)^{D/2-1}.
\label{sol2}
\end{eqnarray}

\begin{figure}[b]
\begin{center}
\includegraphics[width=4.4cm]{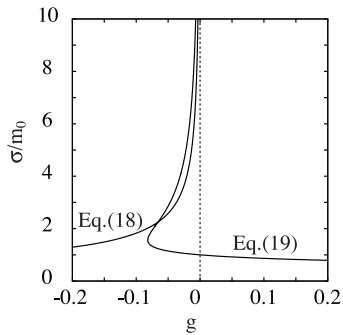}
\caption{Solutions (\ref{sol1}) and (\ref{sol2}) as a function of $g$ at $D=3$.}
\label{f00}
\end{center}
\end{figure}

We plot a typical behavior of the non-trivial solutions (\ref{sol1}) 
and (\ref{sol2}) as a function of $g$ in Fig.~\ref{f00}.
As is shown, the solution (\ref{sol1}) appears for a negative $g$ and disappears at the limit, $G_2 \rightarrow 0$.
In the present paper we assume that the discrete chiral symmetry
is broken for vanishing temperature and chemical potential and
take a negative value for $G_1$. Hence the eight-fermion coupling,
$G_2$, is positive definite for a negative $g$.
The gap equation of the four-fermion interaction model is reproduced at the $g\rightarrow 0$ limit of Eq.~(\ref{sol2}). 

\begin{figure}[tb]
\begin{center}
\includegraphics[width=4.4cm]{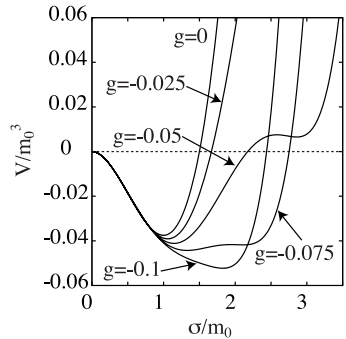}
\caption{Behavior of the effective potential 
for $g=-0.1, -0.075, -0.05,\protect\\ -0.025, 0$.}
\label{f01}
\end{center}
\end{figure}

\begin{figure}[t]
\begin{center}
\includegraphics[width=4.4cm]{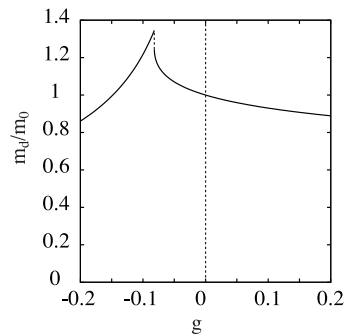}
\caption{Dynamical fermion mass  as a function of $g$ at $D=3$.}
\label{f02}
\end{center}
\end{figure}

It should be noted that the stationary condition is only the necessary condition for the minimum of the effective potential. One of the solutions (\ref{sol0}), (\ref{sol1}) and (\ref{sol2}) gives the expectation value for $\sigma$ on the ground state. The behavior of the effective potential (\ref{v0:2}) is shown in Fig.~\ref{f01} as $g$ varies at $D=3$. We clearly observe a local extremum which corresponds to the solution (\ref{sol2}). It disappears for $g=-0.1$

Evaluating the effective potential, we find the the expectation value, $\langle \sigma \rangle$, on the ground state. From Eq.~(\ref{s1}) the dynamical fermion mass, $m_{d}$, is described as a function of the expectation value, $\langle \sigma \rangle$,
\begin{eqnarray}
\frac{m_{d}}{m_{0}}\equiv \frac{\langle s\rangle}{m_{0}} = 
\frac{\langle \sigma \rangle}{m_{0}}
+ g\left(\frac{\langle \sigma\rangle}{m_{0}}\right)^3 .
\end{eqnarray}
In Fig.~\ref{f02} we plot the dynamical fermion mass at $D=3$. We observe a small mass gap where the global minimum moves between the solutions (\ref{sol1}) and (\ref{sol2}). On the right-hand side of the gap the minimum of the effective potential is derived by Eq.~(\ref{sol2}). For the solution (\ref{sol1}) we find an simple analytical expression for the dynamical fermion mass,
\begin{eqnarray}
\frac{m_{d}}{m_{0}}=\sqrt[]{\mathstrut -\frac{1}{3g}}+g \left( -\frac{1}{3g} \right)^{3/2} .
\label{md:const}
\end{eqnarray}
Thus the dynamical mass is given by Eq.~(\ref{md:const}) on the left-hand side of the mass gap in Fig.~\ref{f02}.

\section{Effective Potential at Finite Temperature and Chemical Potential}
To discuss the phase structure of the model we calculate the effective potential at finite temperature and chemical potential. According to the imaginary time formalism, the spinor two-point function at finite temperature and chemical potential is obtained from Eq. (\ref{Spi2}) by the Wick rotation and the replacements
\begin{eqnarray}
\int_{-\infty}^{\infty} \frac{dk^{0}}{2 \pi} \rightarrow i \frac{1}{\beta} \sum_{n=-\infty}^{\infty} ,
\end{eqnarray}
\begin{eqnarray}
k^{0} \rightarrow i \omega_{n} - \mu \equiv i \frac{2n+1}{\beta}\pi -\mu ,
\end{eqnarray}
\begin{eqnarray}
\gamma^0 \rightarrow \gamma^4\equiv i \gamma^{0} ,
\end{eqnarray}
where $\beta\equiv 1/(k_BT)$ with $k_B$ the Boltzmann constant and $T$ the temperature, and $\mu$ is the chemical potential. Thus the spinor two-point function is modified to be
\begin{eqnarray}
S(x,y;z)=i \frac{1}{\beta} \sum_{n} \int \frac{d^{D-1} {\bf k}}{(2 \pi)^{D-1}}e^{-ik(x-y)} \frac{1}{\gamma^{\mu} k_{\mu}-s} .
\label{Spi2:FTFM}
\end{eqnarray}
Inserting Eq.(\ref{Spi2:FTFM}) into Eq.(\ref{EP}), we obtain the effective potential at finite temperature and chemical potential,
\begin{eqnarray}
V(s,\sigma)&=&-\frac{1}{4G_{1}}\sigma^{2}+\frac{1}{2G_{1}}s \sigma-\frac{G_{2}}{16G_{1}^{4}} \sigma^{4} \nonumber \\
&& -\frac{1}{2\beta}\sum_{n}\int \frac{d^{D-1}{\bf k}}{(2\pi)^{D-1}} \ln \frac{(\omega_{n}+i\mu)^{2}+{\bf k}^{2}+s^{2}}{(\omega_{n}+i \mu)^{2}+{\bf k}^{2}} .
\label{vs0}
\end{eqnarray}
We can perform the summation and integrate over angle variables in this equation. Then we obtain
\begin{equation}
V(s,\sigma)=V_0(s,\sigma) + V^{\beta\mu}(s),
\label{v:FTFM}
\end{equation}
where $V_0(s,\sigma)$ is the effective potential at $T=\mu=0$ and $V^{\beta\mu}(s)$ is given by
\begin{eqnarray}
V^{\beta \mu}(s) &=& -\frac{1}{\beta}\frac{\mathrm{tr}1}{(4\pi)^{(D-1)/2}}\frac{1}{\displaystyle\Gamma\left(\frac{D-1}{2}\right)} 
\nonumber \\
&&  \times \int dk k^{D-2}
\left[ \ln \frac{1+e^{-\beta \ (\sqrt[]{\mathstrut {k}^{2}+s^{2}}+\mu)}}{1+e^{-\beta({k}+\mu)}} + \ln \frac{1+e^{-\beta \ (\sqrt[]{\mathstrut {k}^{2}+s^{2}}-\mu)}}{1+e^{-\beta({k}-\mu)}} \right] .
\nonumber\\
\label{vs}
\end{eqnarray}
It should be noted that the thermal correction, $V^{\beta \mu}(s)$, is positive-definite. Thus the temperature and chemical potential effect suppresses the chiral symmetry breaking.
The numerical uncertainty increases at the zero-temperature limit of Eq.(\ref{vs}). To study the case at $T=0$ we take the limit analytically and use the following expression,
\begin{eqnarray}
V^{\mu}(s) &\equiv& \lim_{\beta \rightarrow \infty }V^{\beta \mu}(s) 
\nonumber \\
&=&-\frac{\mathrm{tr}1}{(4 \pi)^{(D-1)/2}}\frac{1}{\displaystyle\Gamma \left(  \frac{D-1}{2} \right)} \left[ \int_{0}^{\mu} dk k^{D-2}(k-\mu) \right. \nonumber \\
&& \left. \ \ \ \ -\theta\left(\mu-s\right)
\int_{0}^{\sqrt[]{\mathstrut \mu^{2}-s^{2}}}dk k^{D-2}\left( \sqrt[]{\mathstrut k^{2}+s^{2}} - \mu\right)
\right] .
\end{eqnarray}
Integrating over space components in Eq.~(\ref{vs0}), we obtain another expression for the effective potential \cite{IKM},
\begin{eqnarray}
V^{\beta \mu}(s) &=& \frac{\mathrm{tr}1}{(4\pi)^{D/2}D}\Gamma\left(1-\frac{D}{2}\right)(s^2)^{D/2}+
\frac{1}{2\beta}\frac{\mathrm{tr}1}{(4\pi)^{(D-1)/2}}\Gamma\left(\frac{1-D}{2}\right)
\nonumber \\
&&  \times \sum_{n=-\infty}^{\infty}
\left\{
\left[ (\omega_{n}+i\mu)^{2}+s^{2} \right]^{(D-1)/2}
-\left[ (\omega_{n}+i\mu)^{2}\right]^{(D-1)/2}
\right\} .
\nonumber\\
\label{vs2}
\end{eqnarray}
This expression is equivalent to the one in Eq.~(\ref{vs}). We use it for the analytic study of the critical point. 

Since $V^{\beta \mu}(s)$ is independent on the auxiliary field $\sigma$, the stationary condition (\ref{stationary2}) is not modified by the thermal correction. Hence the effective potential (\ref{v:FTFM}) is described as a function of the auxiliary field, $\sigma$, on the stationary condition (\ref{stationary2}),
\begin{eqnarray}
V(\sigma)\equiv V_0(\sigma) + V^{\beta \mu}\left(s = \sigma + g \frac{\sigma^3}{m_0^2} \right).
\label{vsigma}
\end{eqnarray}
Below we evaluate the effective potential (\ref{vsigma}) at finite temperature and chemical potential and study the phase structure of the four- and eight-fermion interaction model.

\section{Phase Structure at Finite Temperature and Chemical Potential}
The broken chiral symmetry is restored at high temperature and chemical potential. In Ref.~\cite{IKM} the phase diagram is illustrated for the four-fermion interaction model in arbitrary dimensions, $2\leq D <4$. Here we extend the analysis to the four- and eight-fermion interaction model.

\subsection{Phase boundary for the second order phase transition}
Before the full analysis of the phase structure, we analytically consider a phase boundary in a restricted case. We assume that the phase transition is of the second order. If the broken chiral symmetry is restored through the second order phase transition, the order parameter, $\langle\sigma\rangle$, continuously disappears at the critical point. The order parameter satisfies the stationary condition for the effective potential (\ref{vsigma}),
\begin{eqnarray}
\frac{\partial V(\sigma)}{\partial \sigma} &=&
\frac{\partial V_0(\sigma)}{\partial \sigma}
\nonumber\\
&& + \frac{\mathrm{tr}1}{(4 \pi)^{(D-1)/2}} \frac{1}{\displaystyle\Gamma \left(  \frac{D-1}{2} \right)} \left(\sigma+g \frac{\sigma^3}{m_{0}^2} \right) \left(1+3g \left( \frac{\sigma}{m_{0}} \right)^{2}\right)
\nonumber \\
&& \times \int dk k^{D-2} \frac{1}{\displaystyle\sqrt[]{\mathstrut k^{2}+\left(\sigma + g \frac{\sigma^3}{m_{0}^2}\right)^{2}}}
\left[\frac{1}{1+ e^{\beta (\sqrt[]{\mathstrut k^{2}+(\sigma + g \sigma^{3}/m_0^2)^{2}}+\mu)}} \right.  
\nonumber \\
&& \left.  \ \ \ \ +\frac{1}{1+ e^{\beta (\sqrt[]{\mathstrut k^{2}+(\sigma + g \sigma^{3}/m_0^2)^{2}}-\mu)}}  \right]
=0.
\label{stcondsigma}
\end{eqnarray}
For $T=\mu=0$ the stationary condition has the solutions (\ref{sol0}), (\ref{sol1}) and (\ref{sol2}). The constant solutions  (\ref{sol0}) and (\ref{sol1}) also satisfy the stationary condition (\ref{stcondsigma}) at finite temperature and chemical potential. The other solution (\ref{sol2}) is modified as
\begin{eqnarray}
1&=& 
\left(1+g\left( \frac{\sigma}{m_{0}} \right)^{2}\right)
\left\{
\left( \left[ \frac{\sigma}{m_{0}} + g \left( \frac{\sigma}{m_{0}} \right)^{3} \right]^{2} \right)^{D/2-1}\right.
\nonumber \\
&& - \frac{2\sqrt[]{\pi}}
 {\displaystyle\Gamma \left(  \frac{D-1}{2} \right)
  \Gamma \left(1-\frac{D}{2} \right)}
  \int dk k^{D-2} \frac{1}{\displaystyle\sqrt[]{\mathstrut k^{2}+\left(\sigma + g \frac{\sigma^3}{m_{0}^2}\right)^{2}}}
\nonumber \\
&& \left. \times \left[\frac{1}{1+ e^{\beta (\sqrt[]{\mathstrut k^{2}+(\sigma + g \sigma^{3}/m_0^2)^{2}}+\mu)}} 
  +\frac{1}{1+ e^{\beta (\sqrt[]{\mathstrut k^{2}+(\sigma + g \sigma^{3}/m_0^2)^{2}}-\mu)}}  \right] \right\}.
\nonumber\\
\label{sol2FTFM}
\end{eqnarray}
For the second order phase transition the order parameter develops continuously  at the critical point. Thus the critical value for the temperature and chemical potential is obtained by taking the $\sigma \rightarrow 0$ limit in Eq.~(\ref{sol2FTFM}),
\begin{eqnarray}
1=\frac{2}{\sqrt[]{\pi}}\frac{\displaystyle \Gamma\left(\frac{3-D}{2}\right)}{\displaystyle \Gamma\left(1-\frac{D}{2}\right)}
 \left(\frac{\beta m_0}{2\pi}\right)^{2-D} \mathrm{Re} \zeta \left(3-D,\frac{1}{2}+i\frac{\beta\mu}{2\pi}\right),
\label{critical:betamu}
\end{eqnarray}
where $\zeta \left(z,a\right)$ is the generalized zeta function. Taking the limit $\mu\rightarrow 0$ in Eq.~(\ref{critical:betamu}) we obtain the critical temperature for $\mu=0$,
\begin{eqnarray}
\beta m_0=2\pi\left[\frac{2}{\sqrt[]{\pi}}\frac{\displaystyle \Gamma\left(\frac{3-D}{2}\right)}{\displaystyle \Gamma\left(1-\frac{D}{2}\right)}
 \left(2^{3-D}-1\right) \zeta \left(3-D\right)\right]^{1/(D-2)} .
\label{critical:beta}
\end{eqnarray}
For $\beta\rightarrow \infty$ Eq.~(\ref{critical:betamu}) simplifies to
\begin{eqnarray}
\frac{\mu}{m_0}=\left[\frac{1}{2}B\left(\frac{4-D}{2},\frac{D-1}{2}\right)\right]^{1/(D-2)} .
\label{critical:mu}
\end{eqnarray}

It is clearly seen that Eqs.~(\ref{critical:betamu}), (\ref{critical:beta}) and (\ref{critical:mu}) are independent on the eight-fermion coupling constant. Therefore the critical values receive no correction from the eight-fermion interaction for the second order phase transition in the leading order of the $1/N$ expansion. The contribution from the irrelevant operator vanishes at the critical point for the second order phase transition. Indeed, Eqs.~(\ref{critical:betamu}), (\ref{critical:beta}) and (\ref{critical:mu}) exactly reproduce the one obtained in the four-fermion interaction model \cite{IKM}.

\subsection{Dynamical fermion mass}
In the broken phase the fermion mass is generated dynamically. In our model the fermion field, $\psi$, acquires a finite mass, if a non-vanishing value is assigned for $\langle \sigma \rangle$. In the symmetric phase the fermion mass term is prohibited. Thus the phase boundary is found by observing the dynamically generated fermion mass. Here we numerically evaluate the effective potential(\ref{vsigma}) and calculate the dynamically generated fermion mass as a function of the temperature and chemical potential.

We want to understand the contribution from the eight-fermion interaction. For this purpose it may be instructive to review the behavior of the dynamically generated fermion mass in the four-fermion interaction model, $g=0$. As is shown in Ref.~\cite{IKM}, the fermion mass vanishes above the critical temperature or chemical potential. The first and the second order phase transition coexist for $2\leq D \leq 3$.  Only the second order phase transition takes place for $3< D <4$. 

\begin{figure}[!htb]
\begin{minipage}{0.49\hsize}
\vspace*{8pt}
\includegraphics[width=4.4cm]{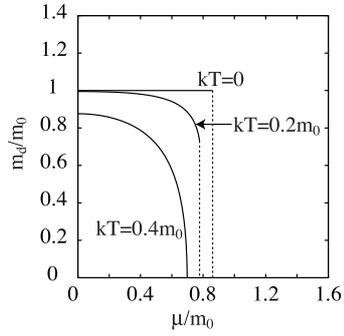}
\begin{center}
(a) $kT/m_{0}$ is fixed.
\end{center}
\end{minipage}
\begin{minipage}{0.49\hsize}
\vspace*{8pt}
\includegraphics[width=4.4cm]{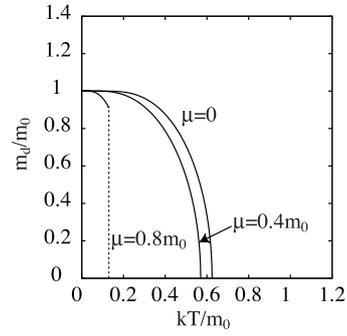}
\begin{center}
(b) $\mu/m_{0}$ is fixed.
\end{center}
\end{minipage}
\caption{Dynamical fermion mass for $D=2.5$ and $g=0$.}
\label{f1}
\end{figure}
\begin{figure}[!htb]
\begin{minipage}{0.49\hsize}
\vspace*{8pt}
\includegraphics[width=4.4cm]{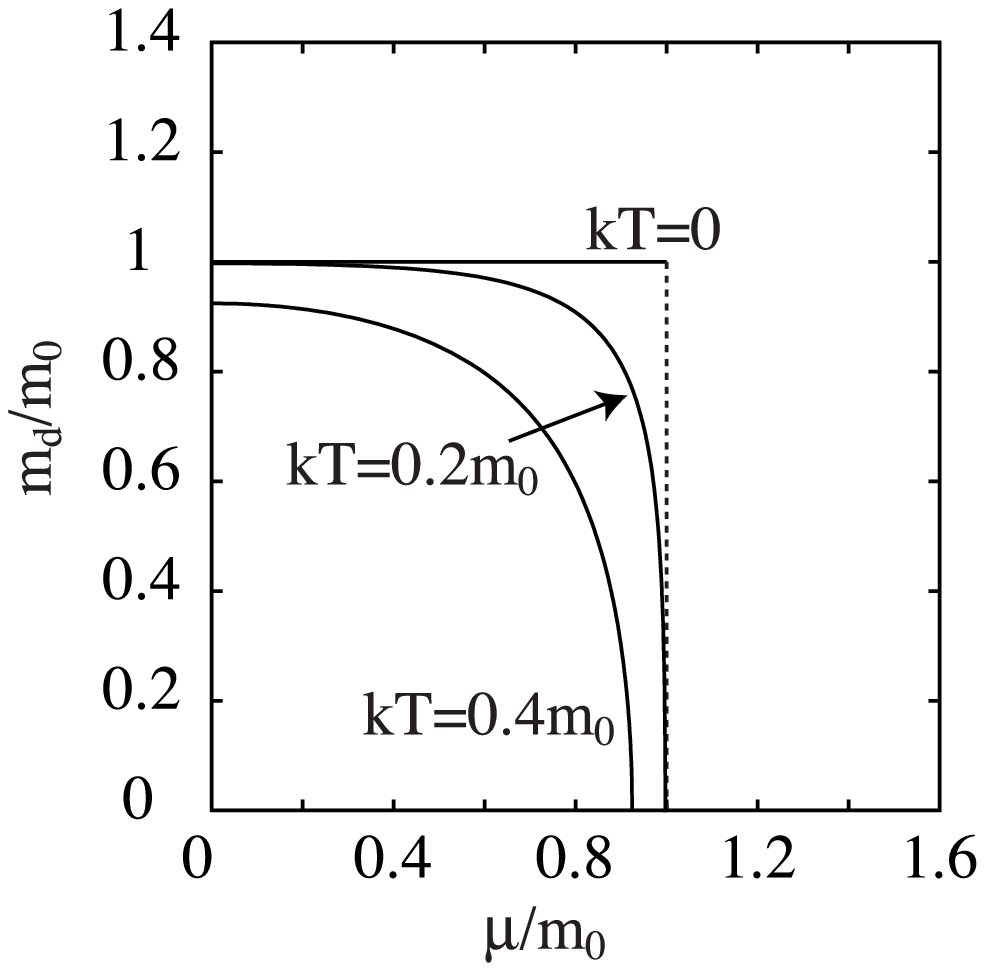}
\begin{center}
(a) $kT/m_{0}$ is fixed.
\end{center}
\end{minipage}
\begin{minipage}{0.49\hsize}
\vspace*{8pt}
\includegraphics[width=4.4cm]{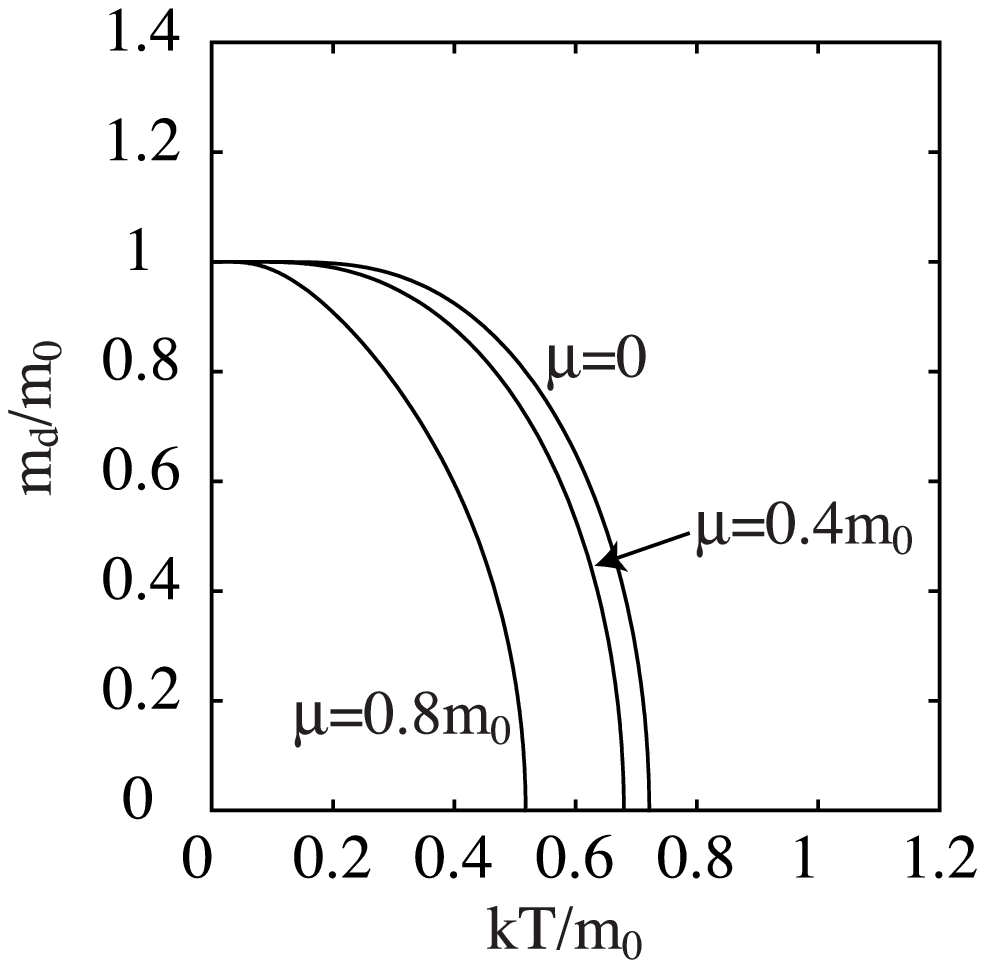}
\begin{center}
(b) $\mu/m_{0}$ is fixed.
\end{center}
\end{minipage}
\caption{Dynamical fermion mass for $D=3.0$ and $g=0$.}
\label{f2}
\end{figure}
\begin{figure}[!htb]
\begin{minipage}{0.49\hsize}
\vspace*{8pt}
\includegraphics[width=4.4cm]{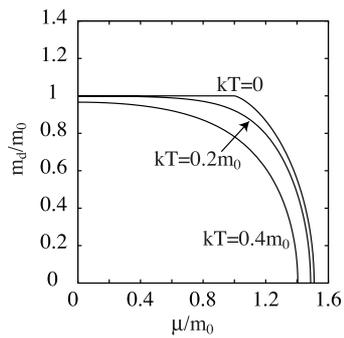}
\begin{center}
(a) $kT/m_{0}$ is fixed.
\end{center}
\end{minipage}
\begin{minipage}{0.49\hsize}
\vspace*{8pt}
\includegraphics[width=4.4cm]{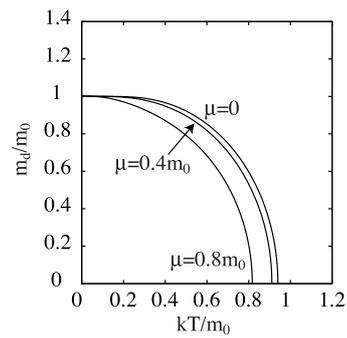}
\begin{center}
(b) $\mu/m_{0}$ is fixed.
\end{center}
\end{minipage}
\caption{Dynamical fermion mass for $D=3.5$ and $g=0$.}
\label{f3}
\end{figure}

In Figs.~\ref{f1}, \ref{f2} and \ref{f3}
we plot the behavior of the dynamical fermion mass for $g=0$. In the case of the first order phase transition a mass gap appears at the critical point. Hence we can clearly distinguish the order of the phase transition. For $D=2.5$ we observe the mass gap for $kT=0, 0.2m_0$ and $\mu=0.8m_0$ in Fig.~\ref{f1} (a) and (b), respectively. The mass gap disappears for higher temperature. There is the first order phase transition at low temperature. In three dimensions the first order phase transition is observed only for $T=0$. There is no first order phase transition for $D=3.5$. In the case of the second order phase transition the critical temperature and chemical potential obey Eq.~(\ref{critical:betamu}).

We introduce the eight-fermion coupling and evaluate the effective 
potential. Since higher dimensional operators that we do not consider 
here have a larger contribution for $\sigma \gg m_0$, it is not valid to 
discuss such a parameter region with only the four- and eight-fermion 
interactions. We do not care about the stability of the effective 
potential and consider the parameter region up to $\sigma \sim$ O$(m_0)$. 
Thus we can set the eight-fermion coupling $g=-0.1$ or $g=0.1$. 

\begin{figure}[!b]
\begin{minipage}{0.49\hsize}
\includegraphics[width=4.4cm]{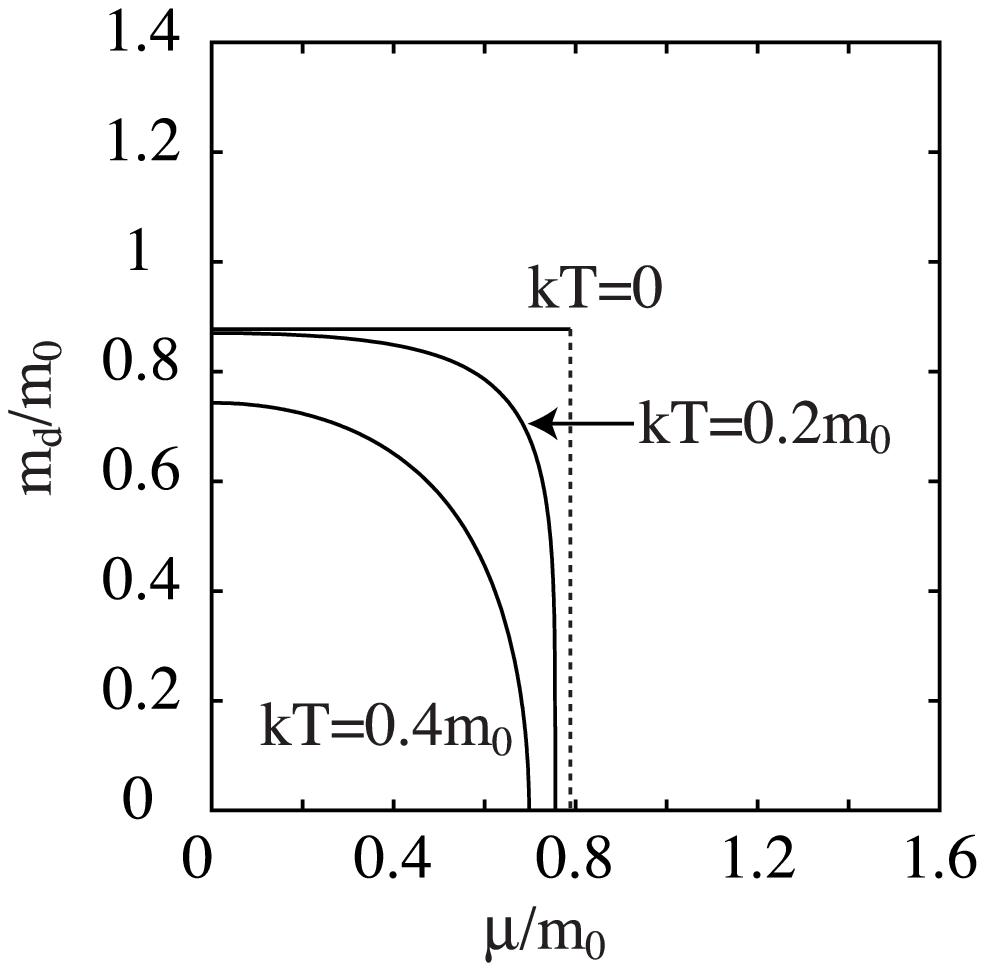}
\begin{center}
(a) $g=0.1.$
\end{center}
\end{minipage}
\begin{minipage}{0.49\hsize}
\includegraphics[width=4.4cm]{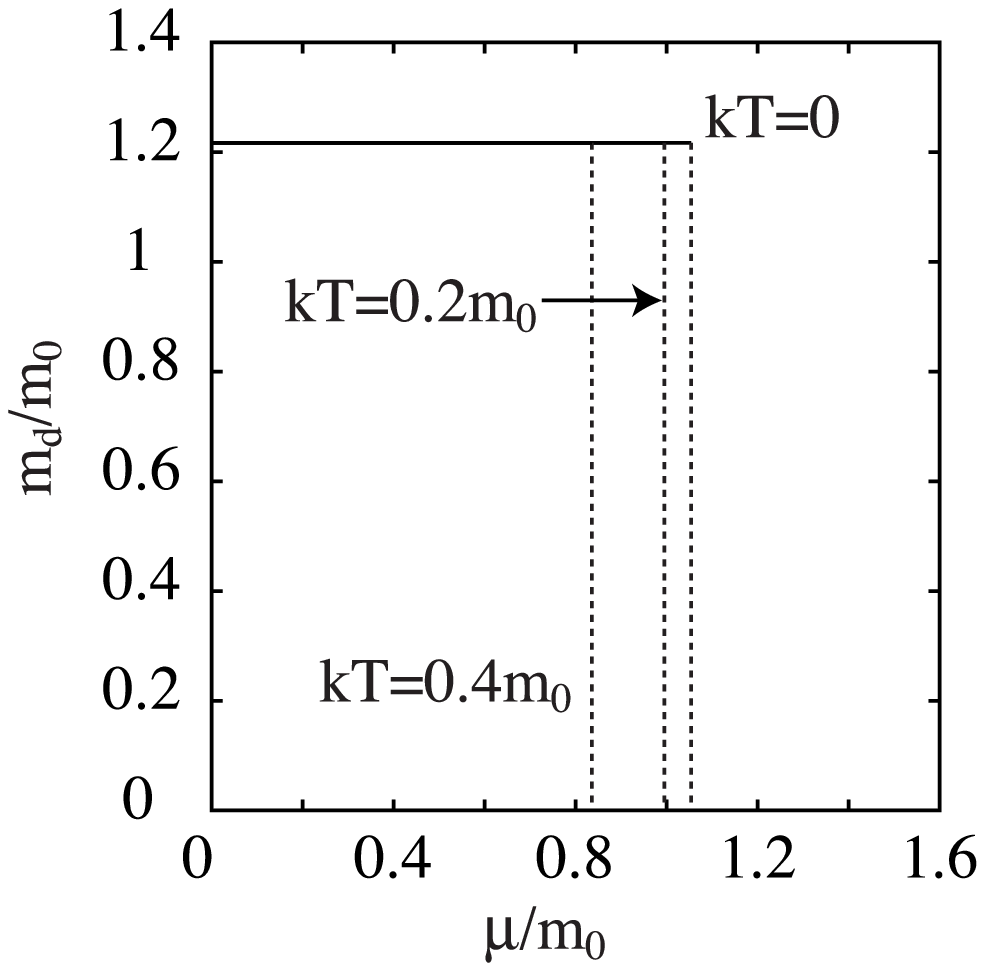}
\begin{center}
(b) $g=-0.1.$
\end{center}
\end{minipage}
\caption{Dynamical fermion mass for $D=2.5$ as a function of the chemical potential $\mu$ with temperature $kT/m_{0}$ fixed at $0, 0.2, 0.4$. }
\label{f7}
\end{figure}

\begin{figure}[t]
\begin{minipage}{0.49\hsize}
\includegraphics[width=4.4cm]{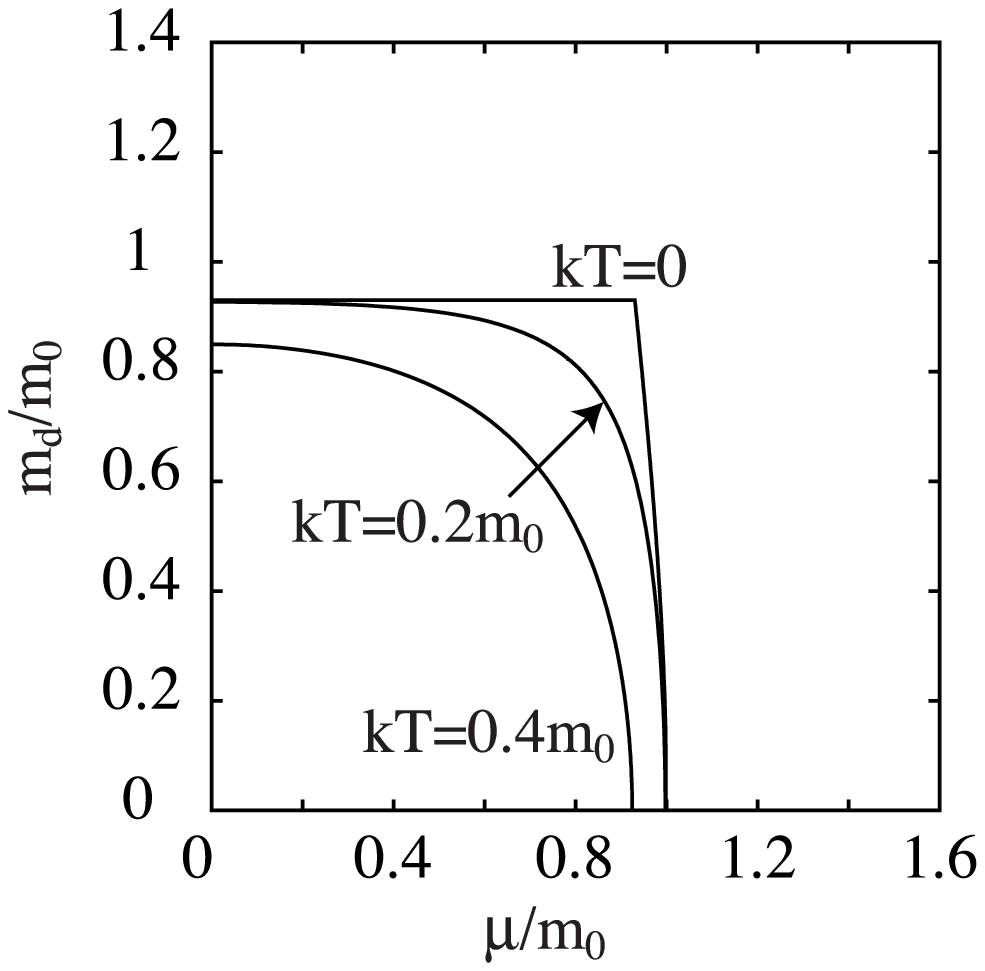}
\begin{center}
(a) $g=0.1.$
\end{center}
\end{minipage}
\begin{minipage}{0.49\hsize}
\includegraphics[width=4.4cm]{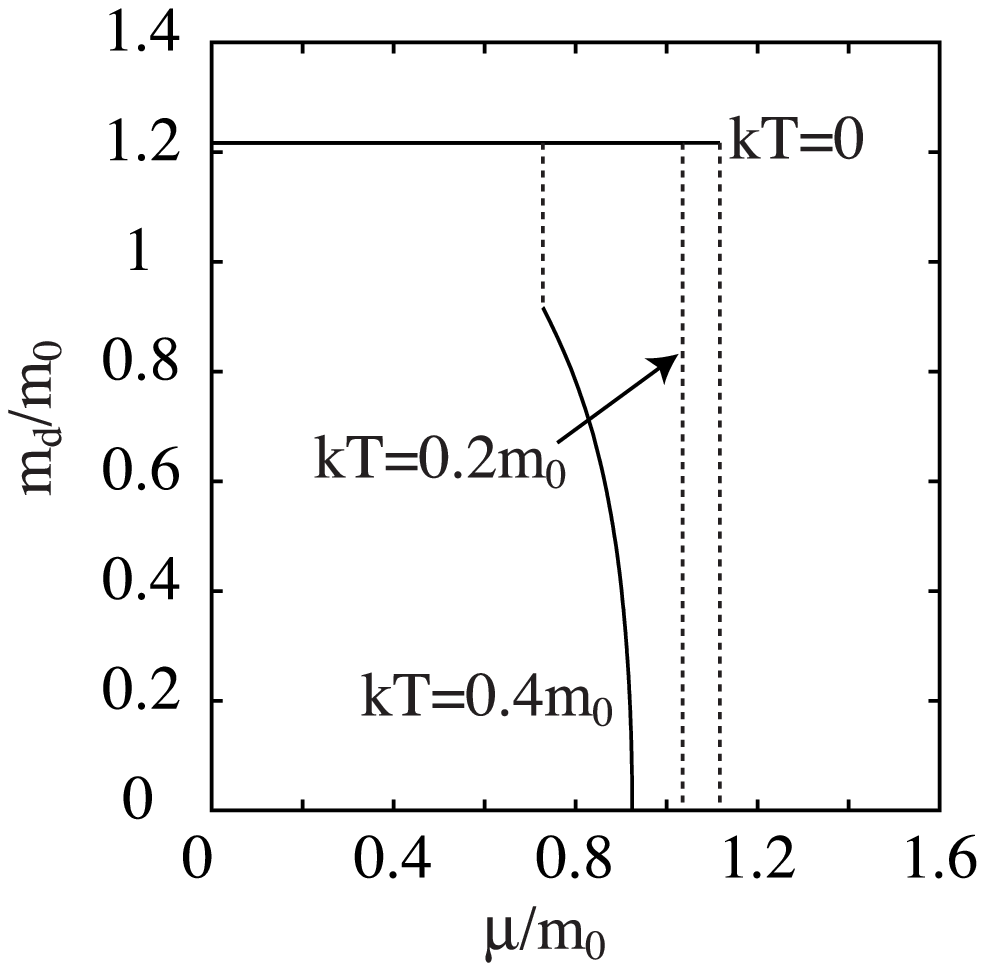}
\begin{center}
(b) $g=-0.1.$
\end{center}
\end{minipage}
\caption{Dynamical fermion mass for $D=3.0$ as a function of the chemical potential $\mu$ with temperature $kT/m_{0}$ fixed at $0, 0.2, 0.4$. }
\label{f8}
\end{figure}

\begin{figure}[t]
\begin{minipage}{0.49\hsize}
\includegraphics[width=4.4cm]{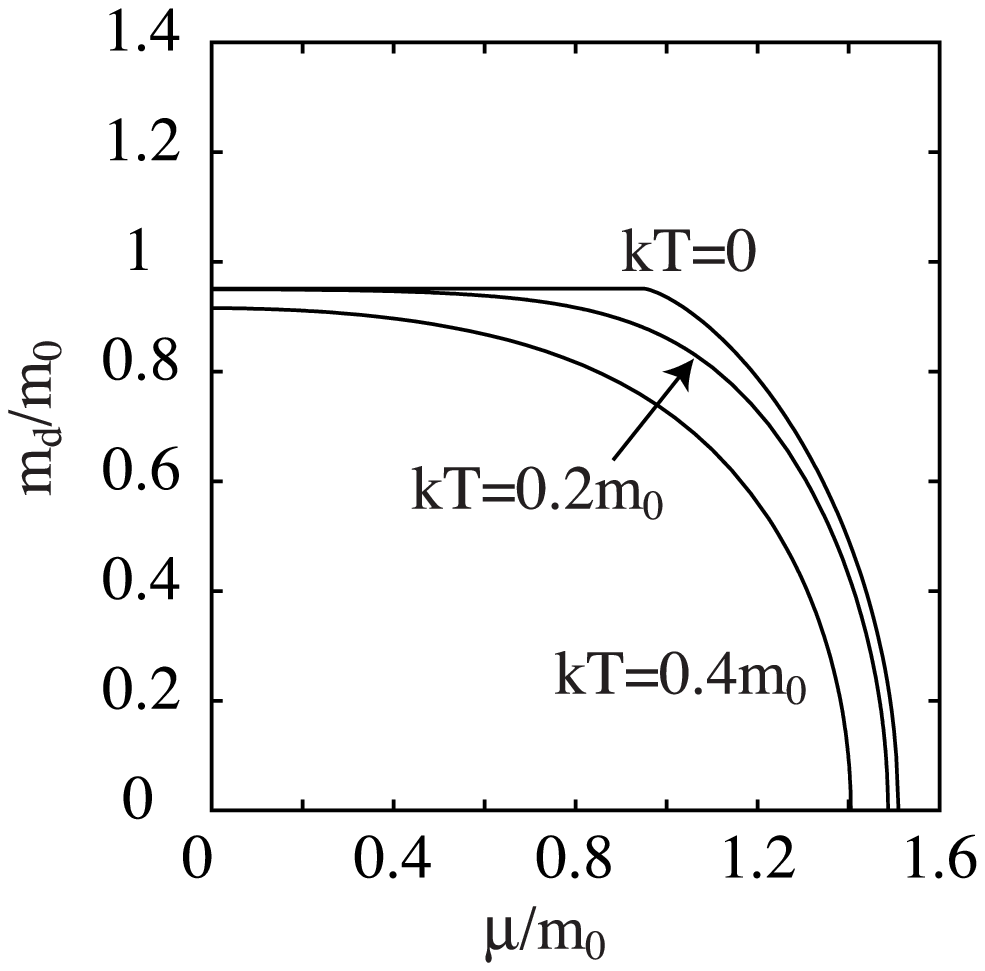}
\begin{center}
(a) $g=0.1.$
\end{center}
\end{minipage}
\begin{minipage}{0.49\hsize}
\includegraphics[width=4.4cm]{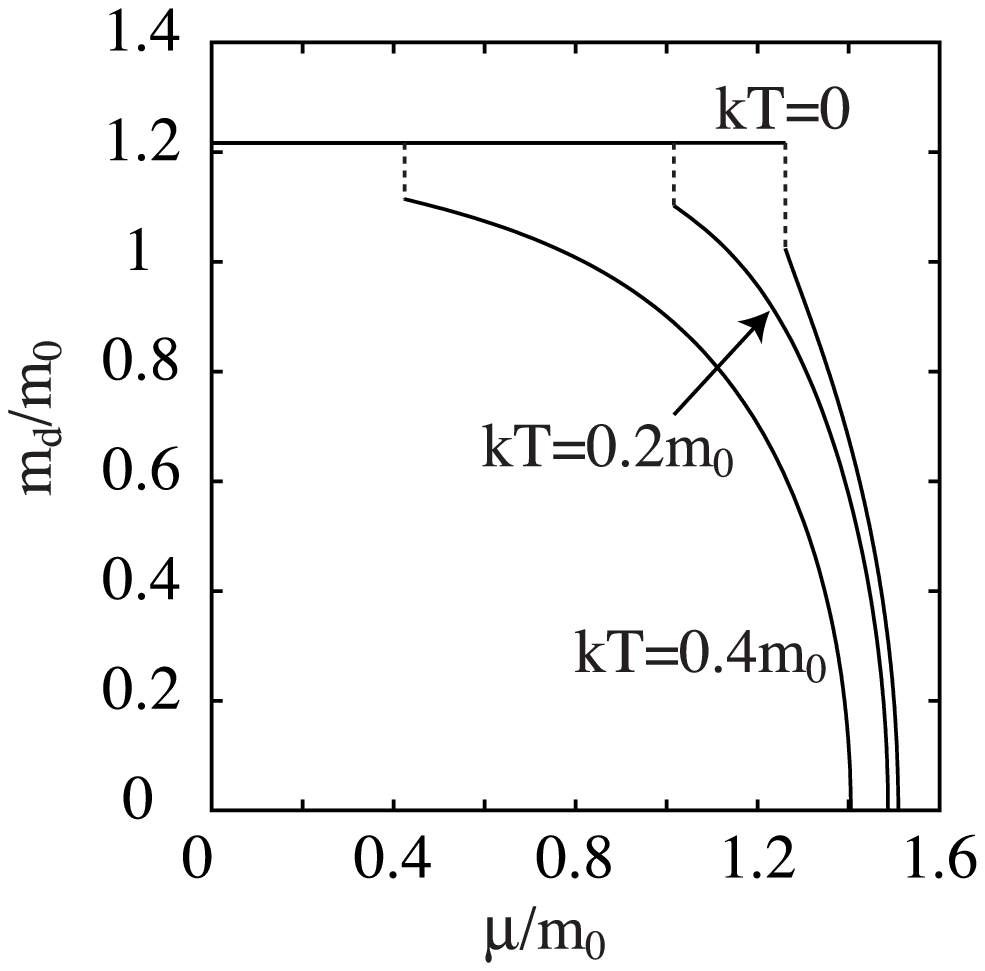}
\begin{center}
(b) $g=-0.1.$
\end{center}
\end{minipage}
\caption{Dynamical fermion mass for $D=3.5$ as a function of the chemical potential $\mu$ with temperature $kT/m_{0}$ fixed at $0, 0.2, 0.4$. }
\label{f9}
\end{figure}

The behavior of the dynamical fermion mass is shown as the chemical 
potential varies with a fixed temperature in Figs.~\ref{f7}, \ref{f8}
and \ref{f9}. A lower and a higher 
fermion mass is generated for a positive and a negative coupling, $g$,
respectively. It is observed that the critical chemical potential 
is not modified by the eight-fermion coupling for the second order 
phase transition. As is shown in Fig.~\ref{f7} (b), there is only 
the first 
order phase transition for a negative coupling, $g=-0.1$, at $D=2.5$.
In three dimensions we observe the first order phase transition at 
$T=0$ for $g=0$. It becomes the second order one in Fig.~\ref{f8} (a).


In Figs.~\ref{f7} (b), \ref{f8} (b) and \ref{f9} (b) the dynamical 
fermion mass 
at $\mu=0$ is corresponds to the one predicted in Eq.(\ref{md:const}). 
Since it is a constant solution for the stationary condition of the 
effective potential, we observe a flat behavior of the dynamical mass. 
The solution moves to the trivial one, $m_d=0$, at the 
critical chemical potential through the first order phase transition
for $D=2.5$ . In Fig.~\ref{f9} (b) we observe a small mass gap where the 
solution (\ref{md:const}) drops to the other one (\ref{sol2}) then
the broken symmetry is restored through the second order phase transition
at the critical chemical potential. In three dimensions we observe both 
the behaviors. The constant solution moves to the trivial one at 
$kT=0$ and $0.2m_0$. A small mass gap is observed for $kT=0.4m_0$.

\begin{figure}[!htb]
\begin{minipage}{0.49\hsize}
\includegraphics[width=4.4cm]{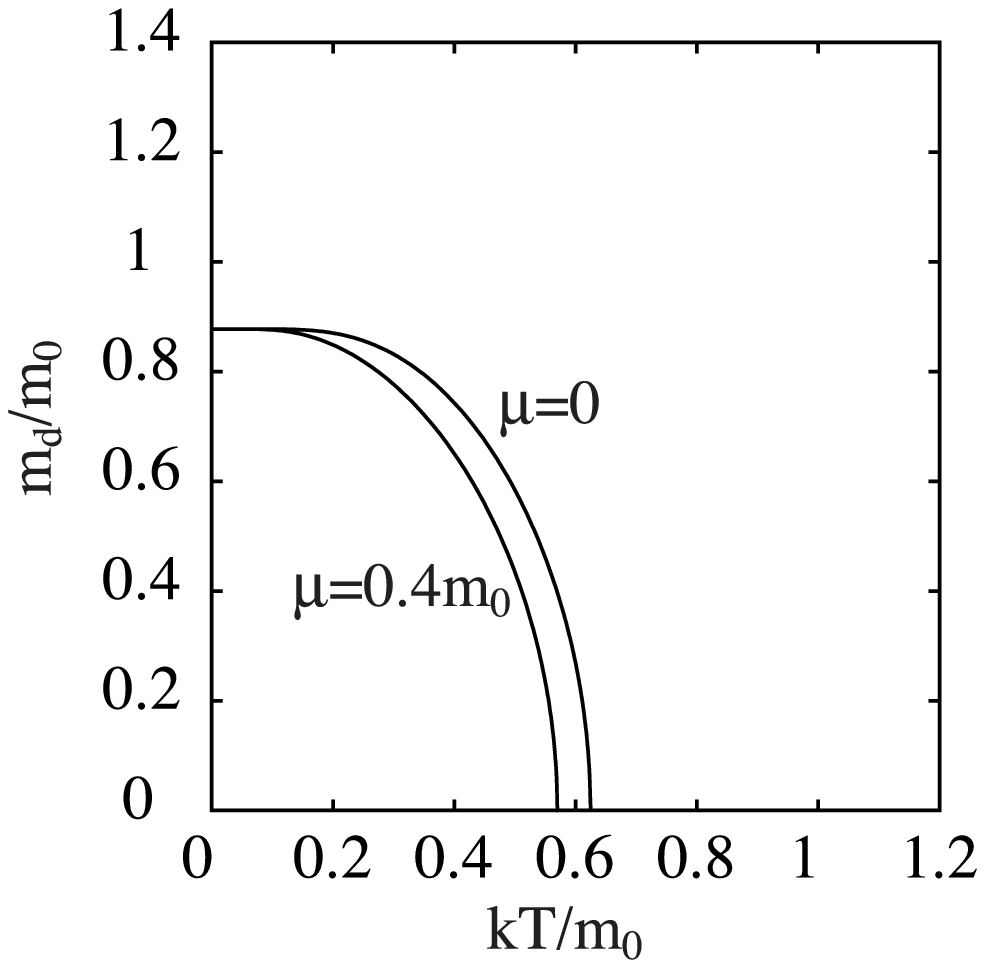}
\begin{center}
(a) $g=0.1.$
\end{center}
\end{minipage}
\begin{minipage}{0.49\hsize}
\includegraphics[width=4.4cm]{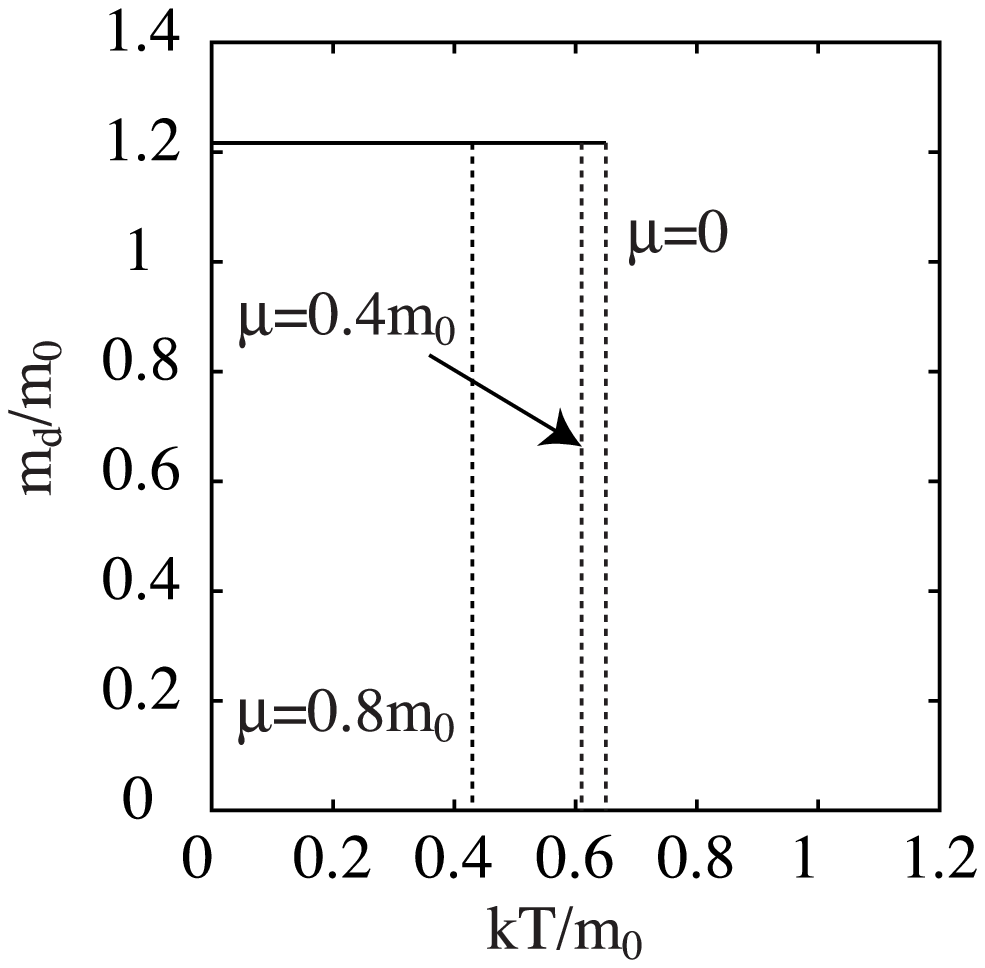}
\begin{center}
(b) $g=-0.1.$
\end{center}
\end{minipage}
\caption{Dynamical fermion mass for $D=2.5$ as a function of the temperature $T$ with chemical potential $\mu/m_{0}$ fixed at $0, 0.4, 0.8$. }
\label{f13}
\end{figure}
\begin{figure}[!htb]
\begin{minipage}{0.49\hsize}
\includegraphics[width=4.4cm]{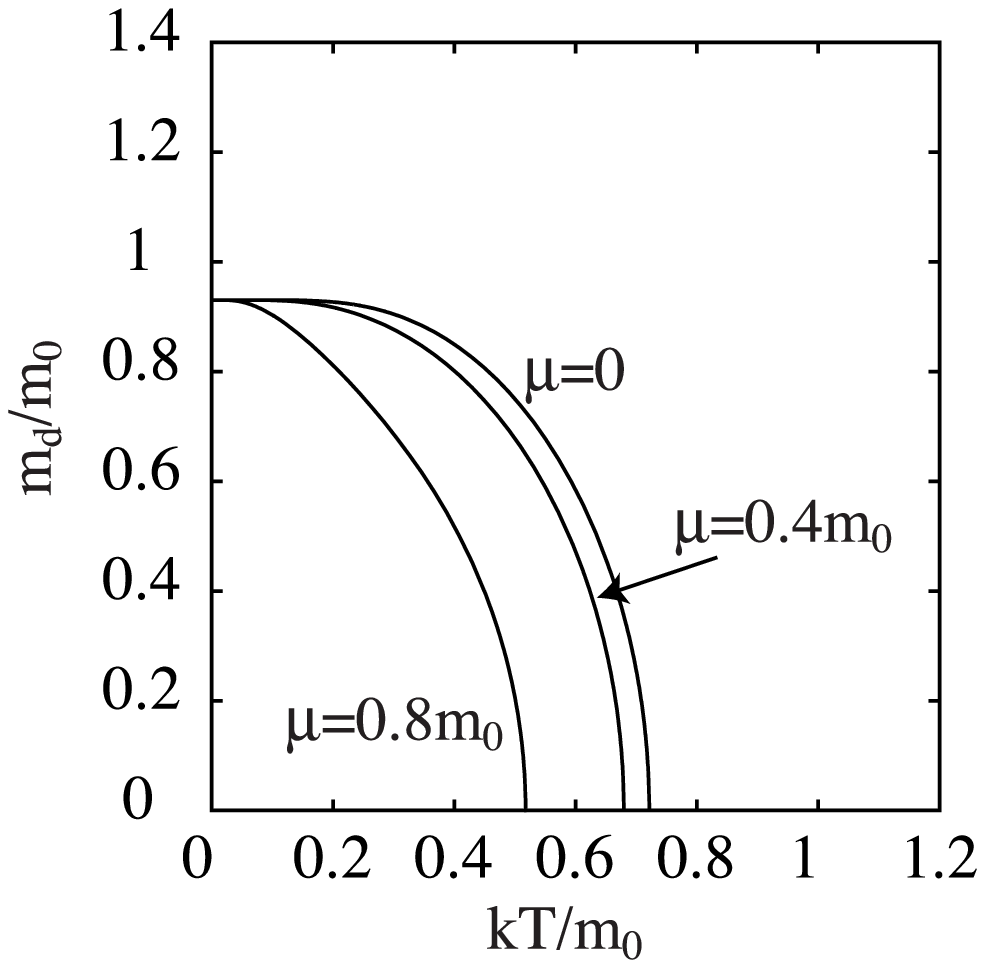}
\begin{center}
(a) $g=0.1.$
\end{center}
\end{minipage}
\begin{minipage}{0.49\hsize}
\includegraphics[width=4.4cm]{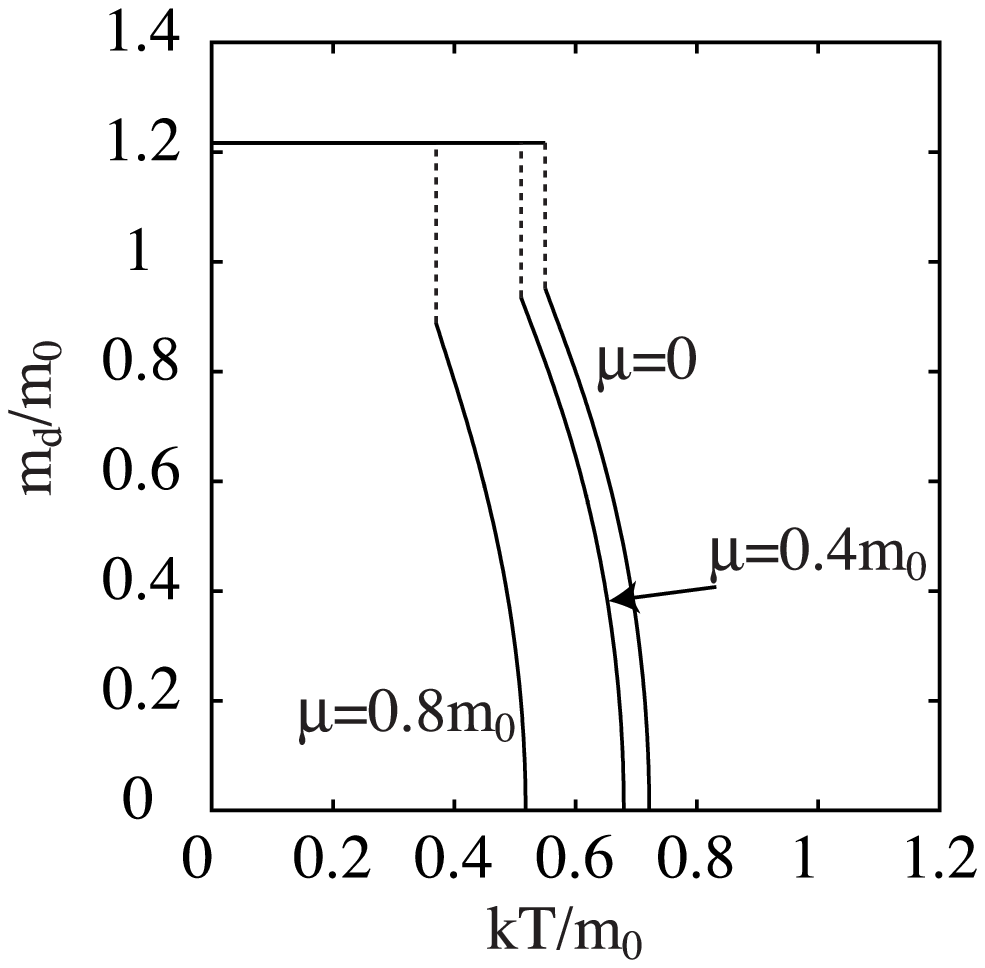}
\begin{center}
(b) $g=-0.1.$
\end{center}
\end{minipage}
\caption{Dynamical fermion mass for $D=3.0$ as a function of the temperature $T$ with chemical potential $\mu/m_{0}$ fixed at $0, 0.4, 0.8$. }
\label{f14}
\end{figure}
\begin{figure}[!htb]
\begin{minipage}{0.49\hsize}
\includegraphics[width=4.4cm]{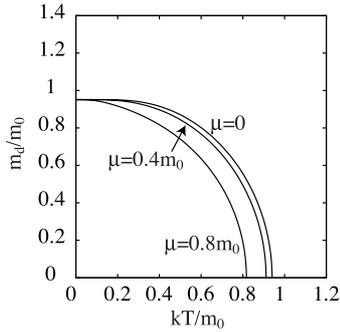}
\begin{center}
(a) $g=0.1.$
\end{center}
\end{minipage}
\begin{minipage}{0.49\hsize}
\includegraphics[width=4.4cm]{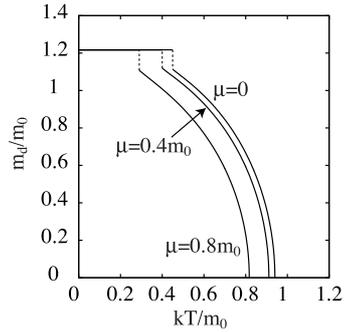}
\begin{center}
(b) $g=-0.1.$
\end{center}
\end{minipage}
\caption{Dynamical fermion mass for $D=3.5$ as a function of the temperature $T$ with chemical potential $\mu/m_{0}$ fixed at $0, 0.4, 0.8$. }
\label{f15}
\end{figure}

In Figs.~\ref{f13}, \ref{f14} and \ref{f15} the behavior of the 
dynamical fermion mass is illustrated
 as a function of the temperature with the chemical potential fixed.
The critical temperature coincides with the one in the four-fermion
interaction model, $g=0$, for the second order phase transition.
We draw only two curves in Fig.~\ref{f13} (a), since the dynamical 
fermion 
mass can not be generated at $\mu=0.8m_0$ for $D=2.5$ and $g=0.1$. 
It means that a positive $g$ suppresses the chiral symmetry
breaking and lowers the critical chemical potential for the first 
order phase transition. The dynamical fermion mass is 
given by Eq.(\ref{md:const}) around $T=0$ in Figs.~\ref{f13} (b), 
\ref{f14} (b) and \ref{f15} (b). We observe a mass gap where 
the solution (\ref{sol1}) moves to the other ones, (\ref{sol0}) 
and (\ref{sol2}).

\begin{figure}[!t]
\begin{minipage}{0.49\hsize}
\includegraphics[width=4.4cm]{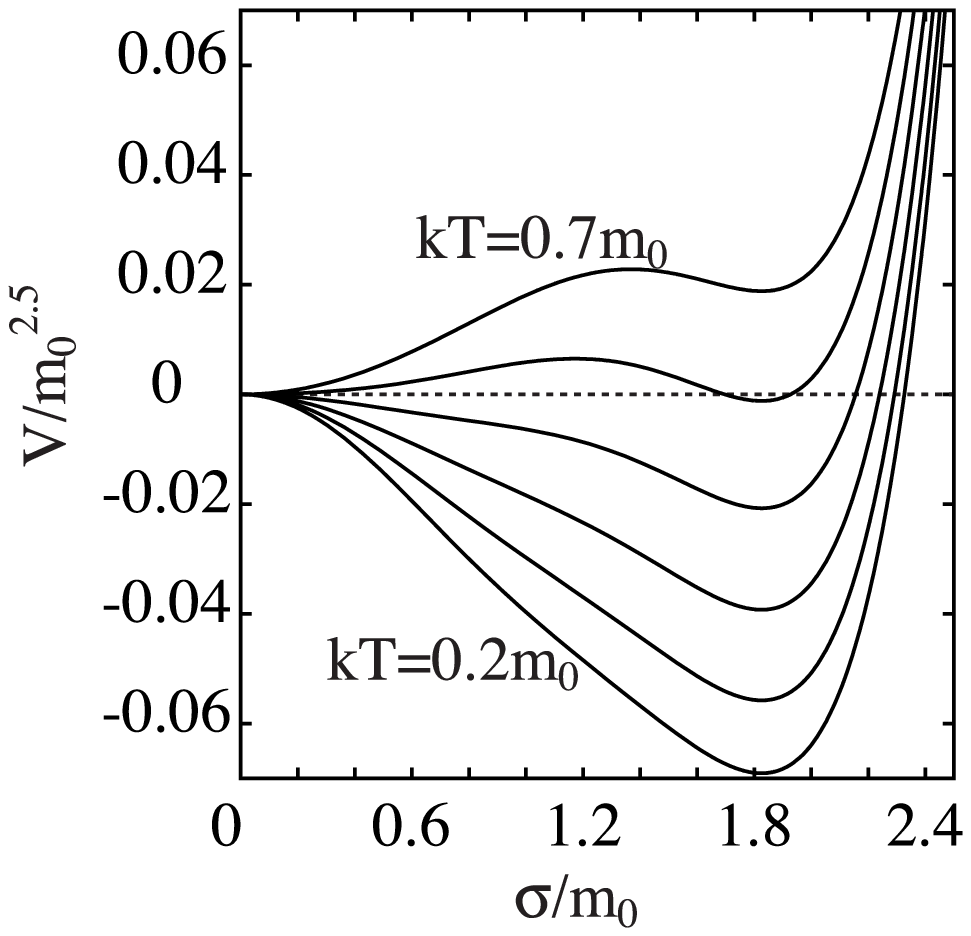}
\begin{center}
(a) $D=2.5$.
\end{center}
\end{minipage}
\begin{minipage}{0.49\hsize}
\includegraphics[width=4.4cm]{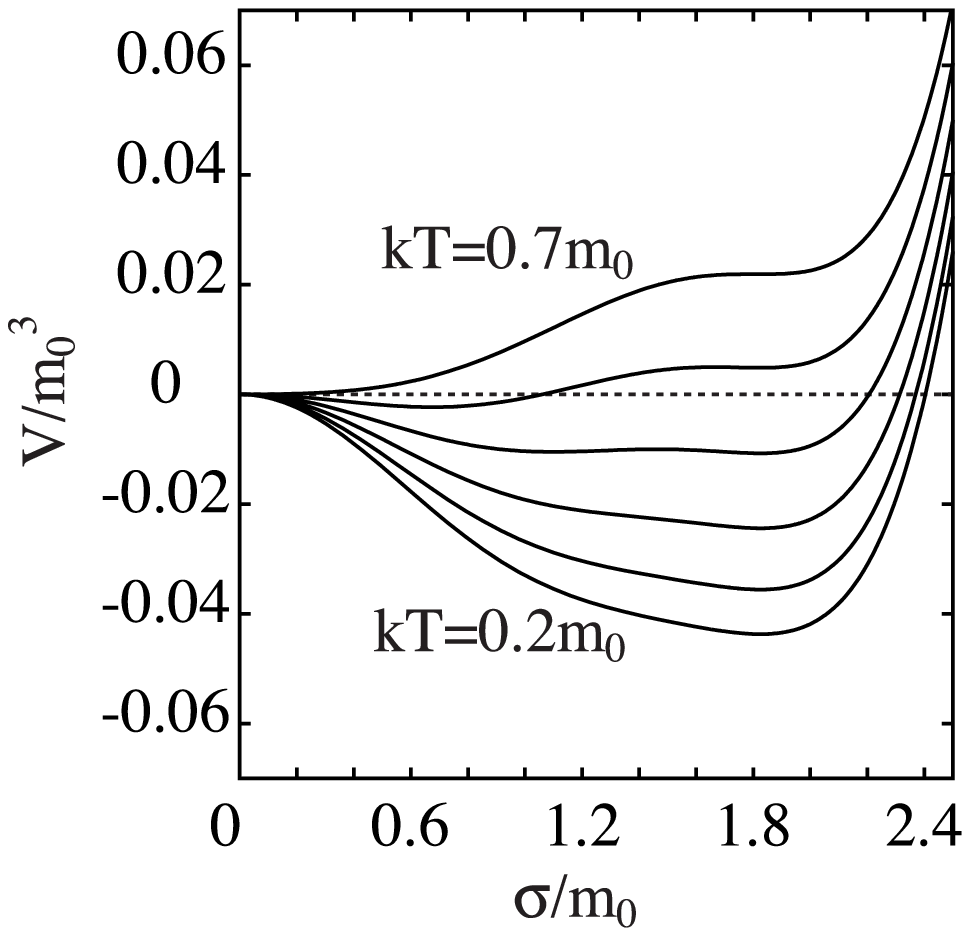}
\begin{center}
(b) $D=3.0$.
\end{center}
\end{minipage}
\caption{Behavior of the effective potential for $g=-0.1$, $kT/m_0=0.2, 0.3, 0.4, 0.5, 0.6, 0.7$ and $\mu/m_{0}=0.4$.}
\label{f22}
\end{figure}

The transition between the solutions (\ref{sol0}), (\ref{sol1}) 
and (\ref{sol2}) is clearly seen by observing the behavior of 
the effective potential. We show a typical one in 
Fig.~\ref{f22}. It should be noticed that the 
effective potential is normalized as $V(0)=0$.
For lower temperature the minimum of the effective potential 
is found at the solution (\ref{sol1}). It moves to the trivial 
one, $\sigma=0$, through the first order phase transition at 
the critical temperature in Fig.~\ref{f22} (a). We observe that 
a local maximum appears between the solution (\ref{sol1}) and 
(\ref{sol2}) in Fig.~\ref{f22} (b). Then the minimum of the 
effective potential continuously approaches to $\sigma=0$. 

\subsection{Phase boundary at finite temperature and chemical potential}
Evaluating the effective potential on the $T-\mu$ plane numerically,
we find the phase boundary which divides the symmetric and the broken 
phase. 
\begin{figure}[!b]
\begin{minipage}{0.49\hsize}
\includegraphics[width=4.4cm]{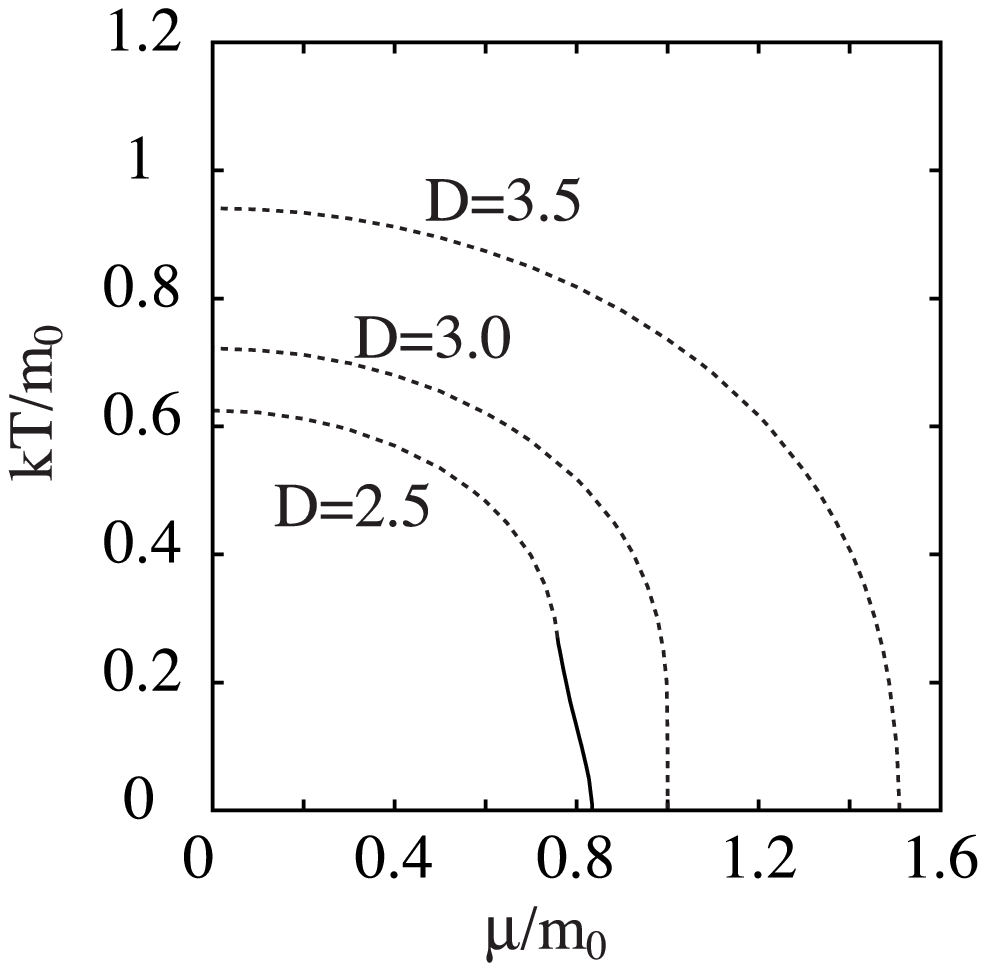}
\caption{Critical lines for $g=0$.}
\label{f24}
\end{minipage}
\begin{minipage}{0.49\hsize}
\includegraphics[width=4.4cm]{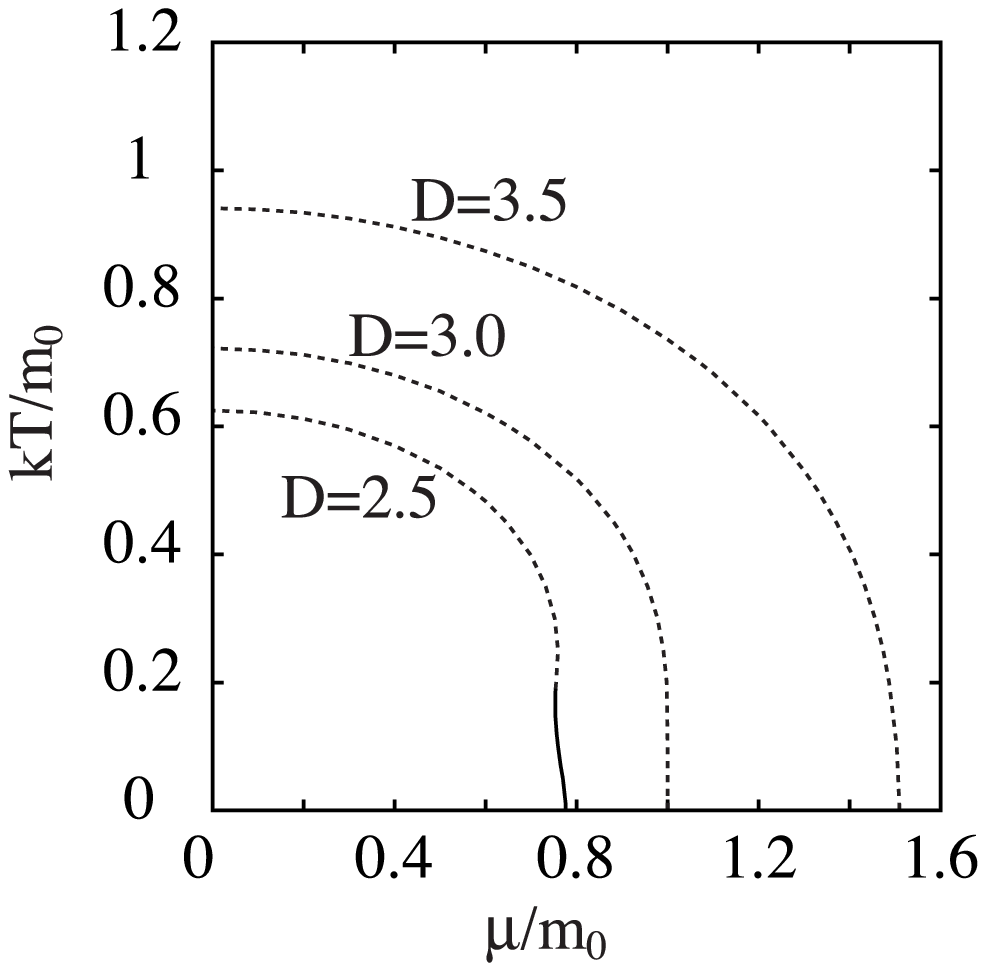}
\caption{Critical lines for $g=0.1$.}
\label{f25}
\end{minipage}
\end{figure}
\begin{figure}[!t]
\begin{center}
\begin{minipage}{0.49\hsize}
\includegraphics[width=4.4cm]{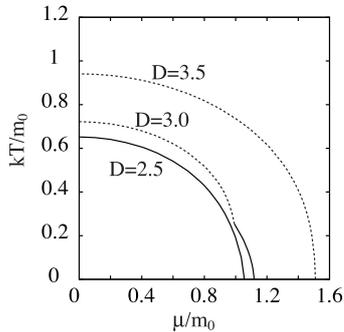}
\caption{Critical lines for $g=-0.1$.}
\label{f26}
\end{minipage}
\end{center}
\end{figure}
We show the phase boundary for the four-fermion interaction model
in Fig.~\ref{f24}. On the dashed lines the broken symmetry is restored 
through the second order phase transition, while the first order phase
transition occurs on the solid lines. We observe the tricritical 
point where the first order critical line meets the second order one
for $2\leq D \leq 3$.

In Figs.~\ref{f25} and \ref{f26} we draw the phase boundary for $g=0.1$
and $-0.1$, respectively. For the second order phase  transition the 
critical temperature and chemical potential have no correction from 
the eight-fermion interaction in the leading order of the $1/N$ 
expansion. Thus we observe that the dashed lines are drawn on the same
place, but some parts are overwritten by the solid lines.
As is shown in Fig.~\ref{f25} the same phase boundary is found
for $D>3$ when $g$ is positive. In $D=2.5$ the tricritical point is 
found at lower temperature. In Fig.~\ref{f26} the first 
order critical line and the tricritical point appear even 
for $D>3$. Only the first order phase transition is found at 
$D=2.5$ for $g=-0.1$.

\section{Conclusion}
We have investigated the phase structure of the four- and 
eight-fermion interaction model at finite temperature and
chemical potential in an arbitrary dimensions, $2\leq D<4$.
Evaluating the effective potential in the leading order of
the $1/N$ expansion, we have calculated the dynamical fermion 
mass and the phase boundary.

It is found that a lighter fermion mass is generated for
a negative eight-fermion coupling, $G_2$, (a positive $g$). 
Because of a constant solution (\ref{sol1}) for the stationary 
condition of the effective potential a new mass gap
from the solution (\ref{sol1}) to (\ref{sol2}) appears for 
a positive $G_2$, (a negative $g$). A similar behavior is
found in Nambu-Jona-Lasinio model with 't Hooft and
eight-quark interactions in four 
dimensions \cite{Moreira:2010in}.

It is analytically shown that there is no correction to 
the second order critical line from the eight-fermion 
interaction. The first order critical line is changed 
and the tricritical point is modified by the 
eight-fermion interaction. Since the four-fermion 
interaction model is renormalizable below 
four-dimensions, only the irrelevant operator in the
model is the eight-fermion interaction. Thus the 
contribution from the irrelevant operator decreases
compared with the renormalizable one near the second 
order critical line and disappears on the line. The
same feature is found for the phase structure in 
a spacetime with non-trivial 
topology \cite{Inagaki:2008zc} and a curved 
spacetime \cite{IH,HI}.

The result seems to be inconsistent with the one in
Refs.~\cite{OHBP} and \cite{Osipov:2006ev}. If 
we consider the model as a regularization of the one 
in four dimensions,
the result depends on the regularization procedure.
It is able to introduce a cut-off scale and define the model
in four dimensions. Since the four- and eight-fermion
interaction are irrelevant, both the operators 
have non-negligible contributes on the second order 
critical line.

At the limit, $\mu\rightarrow 0$, the finite
temperature phase transition is of the second order
in the four-fermion interaction model. Thus the 
same critical temperature is obtained in the model
with a negative $G_2$. The first order phase transition
can be realized for a positive $G_2$. In this case a 
higher critical temperature is obtained.
In the four-fermion interaction model the phase 
transition is of the first order only for $2\leq D\leq 3$
at the limit, $T \rightarrow 0$. Thus the critical
chemical potential takes the same value for a negative 
$G_2$ in $3 <D <4$. For a positive $G_2$ the first order 
phase transition is found even in $3 <D <4$.

In the present paper we only consider four- and eight-fermion
interactions. The eight-fermion interaction produces a new local
minimum for the effective potential and induces the first-order
transition from the local minimum to the other. It is one of the
characteristic features of the contribution from the eight-fermion
interaction. Multi-fermion interactions which given by higher
dimensional operators can also generate additional local minimums.
Then multi-steps of the first order transitions induced in the
phase diagram.

We are interested in applying the results to some physical phenomena.
The first order transition generates a gap for the energy-momentum
tensor for fermion fields. It makes the deceleration parameter of
our universe change suddenly. It also contributes the structure of
dense stars. Thus we expect to observe such effect in cosmological
and astrophysical objects at high temperature and/or high density.
To discuss the physical phenomena we should fix the model parameters
phenomenologically and consider some generalization of the model to
be suitable for each energy scale. To investigate hadron physics
we should include the flavor structure and current quark mass in
the model. The attractive interaction between di-quark channel
plays a decisive role to break the color symmetry at high
density\cite{Dim2}. There is a possibility that super partners
appear at SUSY scale. A supersymmetric extension of the model is
possible along the work in Ref.~\cite{BIO}. It may be important
to study the dynamical symmetry breaking above the SUSY scale.
We continue our work and hope to report on these problems.

\section*{Acknowledgements}
The authors would like to thank D.~Kimura and Y.~Mizutani for fruitful discussions.

\end{document}